\documentclass[journal]{IEEEtran}

\usepackage{amsmath}
\usepackage{amssymb}
\usepackage{graphicx}
\usepackage{tikz}
\usetikzlibrary{positioning,arrows.meta,shapes.geometric,fit,backgrounds,calc}
\usepackage{booktabs}
\usepackage{multirow}
\usepackage{cite}
\usepackage{url}
\usepackage{hyperref}
\usepackage{array}
\usepackage{textcomp}
\usepackage{float}
\usepackage{tabularx}
\usepackage{longtable}

\hypersetup{
    colorlinks=true,
    linkcolor=black,
    citecolor=black,
    urlcolor=blue,
    pdfauthor={Andrii Vakhnovskyi},
    pdftitle={Threat Modeling and Attack Surface Analysis of IoT-Enabled Controlled Environment Agriculture Systems}
}

\begin{document}

\title{Threat Modeling and Attack Surface Analysis of IoT-Enabled Controlled Environment Agriculture Systems}

\author{%
\IEEEauthorblockN{ANDRII VAKHNOVSKYI}\\%
\IEEEauthorblockA{IOGRU LLC, New York, NY 10022, USA}%
\thanks{Corresponding author: Andrii Vakhnovskyi (\mbox{e-mail:} \mbox{andrii.vakhnovskyi@gmail.com}). ORCID: 0009-0007-8306-5932.}%
}

\maketitle

\begin{abstract}
The United States designates Food and Agriculture as one of sixteen critical infrastructure sectors, yet no mandatory cybersecurity requirements exist for agricultural operations, and no formal threat model has been published for Controlled Environment Agriculture (CEA) systems --- greenhouses, vertical farms, and indoor cultivation facilities that depend on tightly coupled IoT sensor networks, industrial control protocols, and cloud-based fleet management. This paper presents the first comprehensive threat model for IoT-enabled CEA, developed through systematic application of STRIDE threat analysis, MITRE ATT\&CK for Industrial Control Systems mapping, and IEC~62443 zone-and-conduit decomposition to a production platform deployed across 30+ commercial facilities in 8~U.S. climate zones. We enumerate 123~unique threats across 25~data-flow-diagram elements spanning 15~communication protocols --- 10 of which (Modbus RTU/TCP, BACnet/IP, 0--10\,V, 4--20\,mA, SDI-12, DALI, I2C, pulse, and dry contact) operate with zero authentication or encryption by design. Threats are scored using DREAD risk assessment and mapped to 19~MITRE ATT\&CK for ICS techniques. We identify five novel attack classes unique to AI-driven CEA: stealth destabilization of neural-network-tuned PID controllers, baseline drift poisoning of anomaly detectors, cross-facility propagation via federated transfer learning, adversarial agronomic schedules that exploit crop biology rather than computational models, and reward poisoning of reinforcement-learning energy optimizers. Physical impact analysis quantifies crop loss timelines: aeroponic systems fail within minutes of irrigation disruption, humidity manipulation triggers pathogen outbreaks within 48--72 hours, and CO\textsubscript{2} injection manipulation creates worker safety hazards at concentrations exceeding NIOSH's Immediately Dangerous to Life or Health (IDLH) threshold. A survey of 10~commercial CEA control vendors reveals that only one CVE has ever been issued against any vendor in this category, zero vendors operate bug bounty programs, and zero hold IEC~62443 cybersecurity certifications. We propose a defense-in-depth countermeasure framework mapped to IEC~62443 security levels and NIST Cybersecurity Framework functions, and recommend that CEA operators target Security Level~2 as a minimum baseline. The complete threat catalog, data flow diagrams, and risk assessment artifacts are published as open-access supplementary materials.
\end{abstract}


\section{Introduction}

\IEEEPARstart{C}{ONTROLLED} Environment Agriculture (CEA) --- encompassing greenhouses, vertical farms, indoor cultivation facilities, and plant factories --- represents a rapidly expanding segment of global food production, with the market projected to exceed \$120~billion by 2030~\cite{ref_cea_market}. CEA facilities depend on dense networks of IoT sensors and industrial actuators to maintain precise environmental conditions: air temperature, relative humidity, CO\textsubscript{2} concentration, photosynthetically active radiation (PAR), substrate moisture, and nutrient solution chemistry must be regulated simultaneously and continuously. Energy expenditure for climate regulation accounts for 20--50\% of CEA operating budgets~\cite{ref_iogru_arxiv}.

Modern CEA platforms integrate three distinct technology tiers: (1)~a field layer of industrial sensors and actuators communicating via legacy protocols including Modbus RTU, BACnet/IP, and analog 4--20\,mA current loops; (2)~an edge computing layer hosting real-time control logic, neural network-based PID auto-tuning, and anomaly detection; and (3)~a cloud layer providing fleet management, cross-facility transfer learning, digital twin simulation, and regulatory compliance integration~\cite{ref_iogru_arxiv}. This three-tier architecture creates a complex attack surface that spans operational technology (OT), information technology (IT), and machine learning (ML) domains.

Despite the critical role of CEA in food security, the cybersecurity of these systems has received remarkably little academic attention. The U.S. Cybersecurity and Infrastructure Security Agency (CISA) designates Food and Agriculture as one of sixteen critical infrastructure sectors~\cite{ref_cisa_sector}, yet compliance with cybersecurity standards remains entirely voluntary. The FBI has issued multiple Private Industry Notifications warning that ransomware actors deliberately time attacks on agricultural cooperatives to coincide with critical planting and harvest seasons~\cite{ref_fbi_pin_2022}. Ransomware attacks on the food and agriculture sector reached 265~incidents in 2025, more than doubling from 2023~\cite{ref_halcyon_2025}. High-profile incidents include the REvil attack on JBS~S.A. that shut down 20\% of U.S. beef processing capacity (ransom: \$11~million)~\cite{ref_jbs}, the BlackMatter attack on NEW Cooperative that threatened 40\% of U.S. grain production software~\cite{ref_new_coop}, and the Everest ransomware breach of STIIIZY that exposed 420,000 cannabis customer records~\cite{ref_stiiizy}.

A systematic literature review reveals a critical gap: while formal threat models exist for smart grids~\cite{ref_jbair_microgrid}, connected vehicles~\cite{ref_automotive_tara}, healthcare IoT~\cite{ref_healthcare_stride}, and building automation systems~\cite{ref_bas_security}, \textbf{no comprehensive threat model has been published for CEA systems}. The two closest works are Fereidooni~\textit{et~al.}~\cite{ref_fereidooni}, who applied STRIDE to precision agriculture (field crops, not CEA) and identified 58~threats, and Tripathi~\textit{et~al.}~\cite{ref_tripathi}, who produced 126~threats for smart greenhouses but with limited depth (only four attack trees) and no coverage of the cloud, ML, or compliance integration layers.

Furthermore, CEA-specific control system vendors --- Priva, Argus Controls, TrolMaster, Wadsworth, Hoogendoorn, Ridder, and Growlink --- have received virtually no public security scrutiny. Our survey (Section~\ref{sec:vendors}) finds that across all major CEA vendors, only one Common Vulnerability and Exposure (CVE) has ever been published, zero vendors operate bug bounty programs, and zero hold IEC~62443 cybersecurity certifications.

This paper addresses these gaps with the following contributions:

\begin{enumerate}
\item We present the first comprehensive threat model for IoT-enabled CEA, applying STRIDE systematically to a three-tier reference architecture deployed across 30+ commercial facilities, enumerating 123~unique threats across 25~data-flow-diagram elements and 15~communication protocols.

\item We map all threats to MITRE ATT\&CK for ICS techniques and score them using DREAD risk assessment, providing a quantitative, prioritized threat catalog.

\item We identify five novel attack classes unique to AI-driven CEA --- including adversarial agronomic schedules, the first reported adversarial ML attack class targeting a biological organism rather than a computational model.

\item We quantify the physical consequences of cyber-physical attacks on CEA systems, including crop loss timelines, worker safety thresholds, and financial impact per attack scenario.

\item We survey 10~commercial CEA control vendors and document the near-total absence of public cybersecurity posture in this industry segment.

\item We propose a defense-in-depth countermeasure framework mapped to IEC~62443-3-3 foundational requirements, NIST Cybersecurity Framework v2.0 functions, and OWASP IoT Top~10~\cite{ref_owasp_iot} categories.
\end{enumerate}

The remainder of this paper is organized as follows. Section~\ref{sec:related} surveys related work. Section~\ref{sec:architecture} describes the CEA reference architecture. Section~\ref{sec:methodology} presents the threat modeling methodology. Section~\ref{sec:threats} enumerates the threat catalog. Section~\ref{sec:impact} quantifies physical impact. Section~\ref{sec:ai_threats} introduces AI/ML-specific threats. Section~\ref{sec:vendors} surveys vendor security posture. Section~\ref{sec:countermeasures} proposes countermeasures. Section~\ref{sec:discussion} discusses implications and limitations. Section~\ref{sec:conclusion} concludes the paper.

\section{Related Work}
\label{sec:related}

\subsection{Threat Modeling Methodologies}

Threat modeling is a systematic approach to identifying, classifying, and prioritizing security threats against a system. The most widely adopted methodology is STRIDE, developed by Microsoft~\cite{ref_shostack}, which categorizes threats into six classes: Spoofing, Tampering, Repudiation, Information Disclosure, Denial of Service, and Elevation of Privilege. STRIDE operates on Data Flow Diagrams (DFDs) and assigns threat categories to each DFD element based on its type (process, data store, data flow, or external entity).

Complementary frameworks include MITRE ATT\&CK for ICS~\cite{ref_attack_ics}, which maps adversary tactics, techniques, and procedures (TTPs) observed in real-world industrial control system attacks to a structured matrix of 12~tactics and over 100~techniques; IEC~62443~\cite{ref_iec62443}, which defines a zones-and-conduits model for industrial network segmentation with four security levels (SL~1--4) based on attacker capability; and DREAD~\cite{ref_dread}, which provides a semi-quantitative risk scoring model across five dimensions: Damage, Reproducibility, Exploitability, Affected users, and Discoverability.

Recent surveys by Xiong and Lagerstr\"om~\cite{ref_xiong_slr} and Shevchenko~\textit{et~al.}~\cite{ref_shevchenko_survey} have cataloged over twelve distinct threat modeling methodologies including PASTA, LINDDUN, OCTAVE, and VAST, each with different strengths. We adopt STRIDE as our primary methodology because of its systematic per-element enumeration, augmented with ATT\&CK for ICS mapping for operational relevance, IEC~62443 zones for countermeasure alignment, and DREAD for quantitative prioritization. This multi-framework approach addresses the common reviewer criticism of over-reliance on a single methodology~\cite{ref_publish_tm}.

\subsection{Cybersecurity in Agriculture}

The academic literature on agricultural cybersecurity has grown rapidly since 2020 but remains focused on precision agriculture (field-level systems: drones, GPS-guided tractors, weather stations) rather than CEA. Key surveys include:

Ferrag~\textit{et~al.}~\cite{ref_ferrag_survey} (2020) surveyed security threats in agricultural IoT, identifying attack vectors across sensor networks, communication channels, and cloud platforms. Gupta~\textit{et~al.}~\cite{ref_gupta_review} (2021) reviewed security of smart farming with a focus on precision agriculture. Two comprehensive 2024 systematic reviews --- one in \textit{Computers and Electronics in Agriculture}~\cite{ref_cea_cyber_2024} analyzing 58~documents and another in \textit{Computers \& Security}~\cite{ref_cs_slr_2024} analyzing 37~articles --- established the state of the art but noted the absence of formal threat models for specific CEA architectures. Kulkarni~\textit{et~al.}~\cite{ref_kulkarni_incidents} compiled the most comprehensive incident catalog, documenting 30~cybersecurity incidents in food and agriculture from 2011--2023.

The emerging discipline of cyber-biosecurity~\cite{ref_murch_2018, ref_duncan_2019} addresses the intersection of biological systems and digital control --- a framing directly applicable to CEA. Duncan~\textit{et~al.}~\cite{ref_duncan_2019} note that agriculture and food account for approximately 20\% of U.S. GDP (\$6.7~trillion) and 15\% of employment (43.3~million jobs), yet cyber-biosecurity protections remain minimal. CEA is uniquely vulnerable because cyberattacks on environmental controls can trigger \textit{biological} consequences --- humidity manipulation induces pathogen outbreaks, nutrient tampering creates plant stress that opens disease windows, and HVAC compromise can push unfiltered air into recirculating hydroponics --- creating cyber-to-biological attack chains that traditional cybersecurity frameworks do not capture.

\subsection{ICS Attack Precedents}

The feasibility of cyber-physical attacks on industrial control systems is well-established. Stuxnet (2010)~\cite{ref_stuxnet} destroyed approximately 1,000 uranium enrichment centrifuges by manipulating variable frequency drive speeds via Siemens S7 PLCs. BlackEnergy (2015)~\cite{ref_blackenergy} targeted the Ukrainian power grid, causing 230,000 people to lose power for six hours. Most relevant to CEA safety systems, TRITON/TRISIS (2017)~\cite{ref_triton} was the first malware designed to attack Safety Instrumented Systems (SIS) --- specifically Schneider Electric Triconex controllers at a Saudi petrochemical plant --- demonstrating that even safety-critical interlocks can be compromised when SIS networks are inadequately segmented. The protocols exploited in these attacks (Modbus, S7comm, OPC) are the same protocols used in CEA systems, making these precedents directly transferable.

The protocols used by CEA systems are also exposed on the public internet at scale. Shodan and Censys scans reveal over 18,700~BACnet devices exposed on UDP port 47808 (60\% in the U.S.), 102,000+~MQTT brokers without authentication on port 1883, and 179~Modbus devices on port 502 across critical infrastructure sectors~\cite{ref_nist_800_82}. While these figures represent all sectors, CEA facilities connected via commercial ISPs with default router configurations are part of this exposed population.

\subsection{Threat Models in Adjacent Domains}

Formal threat models have been published for several domains adjacent to CEA. Jbair~\textit{et~al.}~\cite{ref_jbair_microgrid} applied STRIDE to a smart grid microgrid, demonstrating the case-study validation approach. The building automation system (BAS) security analysis by Kaur~\textit{et~al.}~\cite{ref_bas_security} is particularly relevant because BAS systems share BACnet and Modbus protocols with CEA --- and BAS components have accumulated significant CVE history: the Honeywell/Tridium Niagara Framework alone received 13~critical CVEs in 2025 for root-level remote code execution affecting over 1~million installations globally~\cite{ref_niagara_cves}. The Contemporary Controls BASC-20T BACnet router (CVE-2025-13926) was found to have unauthenticated RCE with the product already obsolete and unpatchable~\cite{ref_basc20t_cve}. These vulnerabilities in BAS middleware and protocol gateways are directly applicable to CEA installations that use the same components. The ICS threat modeling systematic literature review~\cite{ref_ics_tm_slr} (2024) catalogs methodologies applied to industrial control systems but finds no CEA-specific application. The ``Publish Your Threat Models!'' position paper by Kohnfelder and Shostack~\cite{ref_publish_tm} argues that the security benefits of publishing threat models far outweigh the risks, a principle we follow by releasing our complete threat catalog as supplementary material.

\subsection{Gap Statement}

No published work provides a formal, systematic threat model for CEA systems that covers the complete technology stack: field-layer industrial protocols, edge AI controllers, cloud fleet management, compliance integration, and the AI/ML attack surface introduced by neural network PID tuning and cross-facility transfer learning. This paper fills that gap.

\section{CEA Reference Architecture}
\label{sec:architecture}

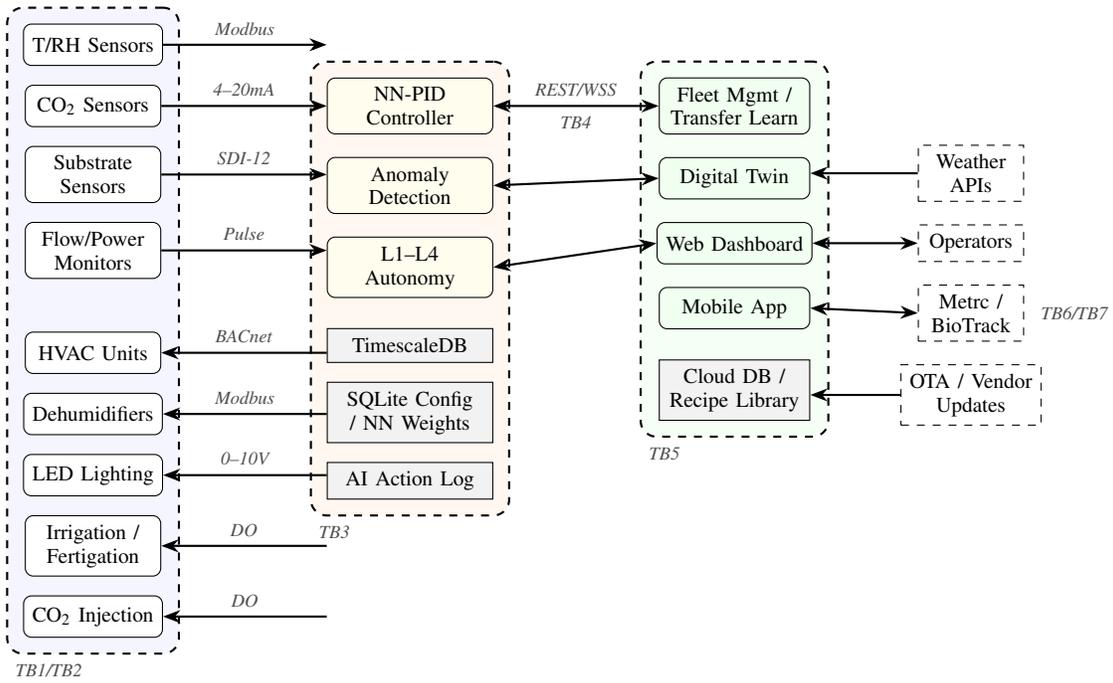
\begin{figure*}[!t]
\centering
\begin{tikzpicture}[
    node distance=0.4cm and 0.6cm,
    >=Stealth,
    every node/.style={font=\footnotesize},
    process/.style={rectangle, draw, rounded corners=3pt, minimum height=0.55cm, minimum width=1.8cm, fill=white, align=center},
    store/.style={rectangle, draw, minimum height=0.45cm, minimum width=1.6cm, fill=gray!10, align=center},
    ext/.style={rectangle, draw, dashed, minimum height=0.45cm, minimum width=1.4cm, fill=white, align=center},
    tb/.style={draw, dashed, thick, rounded corners=6pt, inner sep=6pt},
    flow/.style={->, thick},
    flowbi/.style={<->, thick},
    label/.style={font=\scriptsize\itshape, text=black!70},
]

\node[process] (temp) {T/RH Sensors};
\node[process, below=0.25cm of temp] (co2) {CO\textsubscript{2} Sensors};
\node[process, below=0.25cm of co2] (sub) {Substrate\\Sensors};
\node[process, below=0.25cm of sub] (flow) {Flow/Power\\Monitors};
\node[process, below=0.7cm of flow] (hvac) {HVAC Units};
\node[process, below=0.25cm of hvac] (dehum) {Dehumidifiers};
\node[process, below=0.25cm of dehum] (light) {LED Lighting};
\node[process, below=0.25cm of light] (irr) {Irrigation /\\Fertigation};
\node[process, below=0.25cm of irr] (co2inj) {CO\textsubscript{2} Injection};

\begin{scope}[on background layer]
\node[tb, fill=blue!4, fit=(temp)(co2)(sub)(flow)(hvac)(dehum)(light)(irr)(co2inj), label={[font=\scriptsize\bfseries]above:Field Layer}] (fieldbox) {};
\end{scope}
\node[label, anchor=north west] at (fieldbox.south west) {TB1/TB2};

\node[process, right=2.2cm of co2, minimum width=2.2cm, minimum height=0.7cm, fill=yellow!8] (pid) {NN-PID\\Controller};
\node[process, below=0.3cm of pid, minimum width=2.2cm, fill=yellow!8] (anomaly) {Anomaly\\Detection};
\node[process, below=0.3cm of anomaly, minimum width=2.2cm, fill=yellow!8] (autonomy) {L1--L4\\Autonomy};
\node[store, below=0.4cm of autonomy, minimum width=2.2cm] (tsdb) {TimescaleDB};
\node[store, below=0.25cm of tsdb, minimum width=2.2cm] (sqlite) {SQLite Config\\/ NN Weights};
\node[store, below=0.25cm of sqlite, minimum width=2.2cm] (auditlog) {AI Action Log};

\begin{scope}[on background layer]
\node[tb, fill=orange!6, fit=(pid)(anomaly)(autonomy)(tsdb)(sqlite)(auditlog), label={[font=\scriptsize\bfseries]above:Edge AI Layer}] (edgebox) {};
\end{scope}
\node[label, anchor=north west] at (edgebox.south west) {TB3};

\node[process, right=2.2cm of pid, minimum width=2cm, fill=green!6] (fleet) {Fleet Mgmt /\\Transfer Learn};
\node[process, below=0.3cm of fleet, minimum width=2cm, fill=green!6] (twin) {Digital Twin};
\node[process, below=0.3cm of twin, minimum width=2cm, fill=green!6] (dash) {Web Dashboard};
\node[process, below=0.3cm of dash, minimum width=2cm, fill=green!6] (mobile) {Mobile App};
\node[store, below=0.4cm of mobile, minimum width=2cm] (clouddb) {Cloud DB /\\Recipe Library};

\begin{scope}[on background layer]
\node[tb, fill=green!4, fit=(fleet)(twin)(dash)(mobile)(clouddb), label={[font=\scriptsize\bfseries]above:Cloud Layer}] (cloudbox) {};
\end{scope}
\node[label, anchor=north west] at (cloudbox.south west) {TB5};

\node[ext, right=1.4cm of dash] (user) {Operators};
\node[ext, below=0.3cm of user] (comply) {Metrc /\\BioTrack};
\node[ext, above=0.3cm of user] (weather) {Weather\\APIs};

\node[label, right=0.1cm of comply] {TB6/TB7};

\draw[flow] (temp.east) -- node[above, label] {Modbus} (pid.west|-temp.east);
\draw[flow] (co2.east) -- node[above, label] {4--20mA} (pid.west);
\draw[flow] (sub.east) -- node[above, label] {SDI-12} (anomaly.west|-sub.east);
\draw[flow] (flow.east) -- node[above, label] {Pulse} (anomaly.west|-flow.east);

\draw[flow] (autonomy.west|-hvac.east) -- node[above, label] {BACnet} (hvac.east);
\draw[flow] (autonomy.west|-dehum.east) -- node[above, label] {Modbus} (dehum.east);
\draw[flow] (autonomy.west|-light.east) -- node[above, label] {0--10V} (light.east);
\draw[flow] (autonomy.west|-irr.east) -- node[above, label] {DO} (irr.east);
\draw[flow] (autonomy.west|-co2inj.east) -- node[above, label] {DO} (co2inj.east);

\draw[flowbi] (pid.east) -- node[above, label] {REST/WSS} node[below, label] {TB4} (fleet.west);
\draw[flowbi] (anomaly.east) -- (twin.west);
\draw[flowbi] (autonomy.east) -- (dash.west);

\draw[flowbi] (dash.east) -- (user.west);
\draw[flowbi] (mobile.east) -- (comply.west);
\draw[flow] (weather.west) -- (fleet.east|-weather.west);

\node[ext, below=0.3cm of comply] (ota) {OTA / Vendor\\Updates};
\draw[flow] (ota.west) -- (clouddb.east|-ota.west);

\end{tikzpicture}
\caption{Level~0 Data Flow Diagram of the CEA reference architecture showing three tiers (Field, Edge AI, Cloud), seven trust boundaries (TB1--TB7), and 15~communication protocols. Solid arrows indicate data/command flows; dashed boxes indicate trust boundaries.}
\label{fig:architecture}
\end{figure*}

The threat model is developed against a production CEA IoT platform (IOGRUCloud~\cite{ref_iogru_arxiv}) deployed across 30+ commercial facilities in 8~U.S. climate zones over 7+ years of continuous operation (2017--2024). The architecture comprises three tiers, illustrated in Fig.~\ref{fig:architecture}.

\subsection{Field Layer}

The field layer comprises a distributed network of industrial sensors and actuators communicating via ten OT protocols, none of which provide authentication or encryption by design. Table~\ref{tab:field_protocols} summarizes the protocols and their security properties.

\begin{table}[!t]
\centering
\caption{Field-Layer Communication Protocols}
\label{tab:field_protocols}
\begin{tabular}{@{}llcc@{}}
\toprule
\textbf{Protocol} & \textbf{Use} & \textbf{Auth} & \textbf{Enc.} \\
\midrule
Modbus RTU & T/RH, CO\textsubscript{2}, pH/EC sensors & None & None \\
Modbus TCP & Same, over Ethernet & None & None \\
BACnet/IP & HVAC units (Carrier, Trane) & None & None \\
0--10\,V analog & LED dimming, VFD speed & N/A & N/A \\
4--20\,mA analog & Transmitters (EC, pH, flow) & N/A & N/A \\
SDI-12 & Substrate sensors (TEROS) & None & None \\
DALI & Lighting control & None & None \\
Pulse/dry contact & Flow meters & N/A & N/A \\
I2C & Barometric (internal) & None & None \\
OPC UA & PLC interop (optional) & Cert & TLS \\
\bottomrule
\end{tabular}
\end{table}

Sensor types include aspirated climate stations ($\pm 0.1$\,$^{\circ}$C, $\pm 1.5$\% RH), NDIR CO\textsubscript{2} sensors ($\pm 50$~ppm), quantum PAR sensors, capacitance substrate probes (VWC, EC, temperature), inline pH/EC transmitters, pulse flow meters, differential pressure transmitters, CT clamp power monitors, and IR leaf temperature sensors. Actuators include HVAC units from five major manufacturers (Carrier, Trane, Daikin, LG, Mitsubishi) controlled via BACnet/IP, dehumidifiers (Quest, Anden) via Modbus, LED lighting (Fluence, Gavita, Phantom) via 0--10\,V or DALI, irrigation valves via digital output, fertigation dosing pumps via Modbus, and CO\textsubscript{2} injection solenoids via digital output. A typical zone deploys 20--40 sensors and 8--15 actuators.

\subsection{Edge AI Layer}

The edge layer consists of industrial controllers hosting real-time control logic. Each controller operates autonomously without cloud dependency (air-gapped failsafe) and includes:

\begin{itemize}\item TimescaleDB for time-series sensor storage (10-second resolution, 1-second for anomaly windows)
\item SQLite for configuration: recipes, setpoints, automation rules, I/O mappings, BACnet object tables
\item A 7-3-3 multi-layer perceptron (MLP) for neural network PID auto-tuning with Lyapunov stability guarantees
\item Autoencoder-based anomaly detection on sensor data streams
\item A four-level progressive autonomy model (L1: monitoring, L2: recommendations, L3: autonomous with guardrails, L4: full optimization)
\item Hardware watchdog timer for safety reset
\end{itemize}

Communication to the cloud layer uses REST API and WebSocket over TLS. Communication to field devices uses the OT protocols in Table~\ref{tab:field_protocols}.

\subsection{Cloud Layer}

The cloud layer provides fleet-wide services:

\begin{itemize}\item Aggregated telemetry storage from 30+ facilities
\item Cross-facility transfer learning: VPD trajectories, PID parameters, seasonal templates, and anomaly baselines are learned at one facility and deployed to others
\item Digital twin simulation for recipe testing
\item Web dashboard and mobile application for remote monitoring and control
\item Compliance integration with regulatory tracking systems (Metrc, BioTrack, LeafLogix, Dutchie) via REST API and webhooks
\item Weather forecast ingestion for predictive pre-cooling
\end{itemize}

\subsection{Trust Boundaries}

We identify seven trust boundaries (TB1--TB7) for STRIDE analysis:

\begin{enumerate}\item \textbf{TB1 --- Physical Facility Perimeter}: physical access to sensors, actuators, wiring, and controllers.
\item \textbf{TB2 --- OT Network Boundary}: the industrial control network carrying Modbus RTU, BACnet/IP, analog, and serial protocols --- all unauthenticated.
\item \textbf{TB3 --- Edge Controller Boundary}: bridges OT protocols to IP-based cloud communication; hosts all control logic; the single most critical asset.
\item \textbf{TB4 --- IT/Internet Boundary}: edge-to-cloud communication over TLS (REST, WebSocket, MQTT).
\item \textbf{TB5 --- Cloud Platform Boundary}: fleet data, user accounts, digital twin, and ML models.
\item \textbf{TB6 --- Third-Party Integration Boundary}: compliance platforms, weather APIs, equipment vendor APIs.
\item \textbf{TB7 --- User/Operator Boundary}: web dashboard and mobile application with role-based access.
\end{enumerate}

\section{Threat Modeling Methodology}
\label{sec:methodology}

\subsection{Multi-Framework Approach}

We apply a composite methodology combining four established frameworks:

\textbf{STRIDE}~\cite{ref_shostack} provides the primary threat enumeration engine. We decompose the architecture into 25~DFD elements (10~sensor types, 5~actuator types, 1~edge controller, 4~communication channels, 3~cloud components, and 2~external integrations) and apply the six STRIDE categories to each element systematically.

\textbf{MITRE ATT\&CK for ICS}~\cite{ref_attack_ics} maps each enumerated threat to real-world adversary techniques observed in industrial environments, grounding our analysis in operational threat intelligence rather than purely theoretical risks.

\textbf{IEC~62443}~\cite{ref_iec62443} provides the zone-and-conduit decomposition for organizing countermeasures and defining target security levels. We map the seven trust boundaries to five IEC~62443 zones (Zone~0: field sensors/actuators at SL~1--2; Zone~1: edge controllers at SL~2--3; Zone~2: supervisory/HMI at SL~2--3; Zone~3: site operations at SL~3; Zone~4: enterprise/cloud at SL~3--4).

\textbf{DREAD}~\cite{ref_dread} provides semi-quantitative risk scoring. Each threat receives a score from 1--10 on five dimensions (Damage, Reproducibility, Exploitability, Affected users, Discoverability), yielding a composite score from 5--50.

\subsection{Attacker Model}

We consider four attacker profiles:

\begin{enumerate}\item \textbf{Remote External Attacker}: accesses the system via exposed cloud APIs, MQTT brokers, or BACnet/Modbus services reachable from the internet. Capability: moderate (script kiddie to intermediate). Motivation: financial (ransomware), competitive espionage.
\item \textbf{Insider/Technician}: has legitimate physical or logical access to the facility. Capability: moderate to high. Motivation: sabotage, IP theft, financial gain.
\item \textbf{Supply Chain Attacker}: compromises a CEA equipment vendor, firmware update channel, or ML model repository. Capability: high. Motivation: espionage (nation-state), pre-positioning.
\item \textbf{Nation-State Actor}: targets CEA as critical infrastructure for geopolitical disruption or agricultural IP theft. Capability: very high. Motivation: food security disruption, economic warfare.
\end{enumerate}

The Food and Ag-ISAC has identified 72~active threat actors targeting farm-to-table supply chains~\cite{ref_food_ag_isac}, and Hunt \& Hackett documented 111~APT groups across agriculture, biotech, and industrial sectors~\cite{ref_hunt_hackett}. Russia accounts for 59.3\% of observed adversary activity against the agriculture sector, followed by China at 25.4\%. Agricultural IP theft has been prosecuted at the federal level: in \textit{United States v. Mo Hailong} (2016)~\cite{ref_mo_hailong}, a Chinese national orchestrated a five-year conspiracy to steal proprietary inbred corn seed from DuPont Pioneer and Monsanto, resulting in \$30M+ in losses and 5--8 years of R\&D theft. In \textit{United States v. Haitao Xiang} (2017)~\cite{ref_haitao_xiang}, a Monsanto researcher stole the ``Nutrient Optimizer'' predictive algorithm on a micro SD card and was intercepted at O'Hare Airport en route to China. These cases demonstrate that agricultural automation IP --- including the control algorithms and crop recipes managed by CEA platforms --- is an established target of state-sponsored economic espionage.

\subsection{Scope and Limitations}

The threat model covers the complete CEA IoT stack from field sensors through cloud services. It does not cover physical security of the building envelope (locks, fences), supply chain integrity of seed/genetic material, or business process threats (financial fraud, regulatory capture). The model is validated against a single vendor's platform (IOGRUCloud); generalizability to other CEA architectures is discussed in Section~\ref{sec:discussion}.

\section{Threat Catalog}
\label{sec:threats}

Systematic STRIDE analysis across 25~DFD elements yields 123~unique threats. Table~\ref{tab:stride_summary} summarizes the distribution by STRIDE category and DFD element group.

\begin{table}[!t]
\centering
\caption{Threat Distribution by STRIDE Category}
\label{tab:stride_summary}
\begin{tabular}{@{}lc@{}}
\toprule
\textbf{STRIDE Category} & \textbf{Count} \\
\midrule
Spoofing (S) & 22 \\
Tampering (T) & 28 \\
Repudiation (R) & 15 \\
Information Disclosure (I) & 18 \\
Denial of Service (D) & 19 \\
Elevation of Privilege (E) & 21 \\
\midrule
\textbf{Total} & \textbf{123} \\
\bottomrule
\end{tabular}
\end{table}

The complete threat catalog is provided as supplementary material. Here we present representative threats from each architectural tier, selected by severity (DREAD~$\geq$~40).

\subsection{Field Layer Threats}

The field layer presents the largest attack surface due to the complete absence of authentication and encryption in nine of ten OT protocols. Representative high-severity threats include:

\textbf{T007 --- CO\textsubscript{2} Sensor Spoofing (S, DREAD: 44):} A rogue 4--20\,mA current source injected in parallel with a legitimate CO\textsubscript{2} NDIR sensor reports falsely low CO\textsubscript{2} concentrations. The edge controller responds by increasing CO\textsubscript{2} injection, potentially raising levels above OSHA's 5,000~ppm Permissible Exposure Limit or NIOSH's 40,000~ppm Immediately Dangerous to Life or Health (IDLH) threshold. Attack hardware cost: approximately \$30 for a precision current source. Mapped to MITRE ATT\&CK T0848 (Rogue Master/Relay).

\textbf{T030 --- HVAC Setpoint Tampering (T, DREAD: 46):} BACnet WriteProperty service is used to modify the Analog Value object representing the temperature setpoint on a chiller or rooftop unit. BACnet/IP provides no authentication for write operations; any device on the BACnet VLAN can issue a WriteProperty command. Setting the setpoint to 40$^{\circ}$C causes complete crop loss in flowering-stage cannabis within 4--8 hours. Mapped to T0836 (Modify Parameter).

\textbf{T041 --- Lighting Dark-Period Violation (T, DREAD: 44):} A DALI Direct Arc Power Control (DAPC) broadcast command forces all grow lights to maximum output during the required 12-hour dark period. For short-day plants such as cannabis, even brief light interruption during the dark period prevents flowering and can trigger hermaphroditism --- pollen sac development that converts seedless flower valued at \$1,500--3,000/lb to seeded product worth \$100--200/lb, a 90--95\% value destruction. Mapped to T0836 (Modify Parameter).

\textbf{T055 --- CO\textsubscript{2} Injection Override (T, DREAD: 48):} The digital output controlling the CO\textsubscript{2} solenoid is latched in the energized state via GPIO manipulation, maintaining injection beyond the setpoint while simultaneously suppressing the CO\textsubscript{2} high alarm. In a sealed 10,000~cu~ft grow room with a 50~SCFH injection system, CO\textsubscript{2} can reach IDLH concentrations within hours. CO\textsubscript{2} is odorless; workers cannot detect rising levels without dedicated monitors. This is the highest-severity threat in the catalog: it combines cyber-physical manipulation with direct life-safety consequences. Mapped to T0836 (Modify Parameter).

\subsection{Edge Layer Threats}

The edge controller is the single most critical asset, bridging unauthenticated OT protocols to the cloud:

\textbf{T064 --- PID Loop Denial of Service (D, DREAD: 42):} A fork bomb or out-of-memory condition on the edge controller halts the PID control loop cycle. Without active climate management, room conditions drift to ambient --- potentially exceeding 35$^{\circ}$C in summer or dropping below 10$^{\circ}$C in winter --- causing crop damage proportional to the duration of the outage. Mapped to T0814 (Denial of Service).

\textbf{T066 --- Autonomy Level Escalation (E, DREAD: 44):} The progressive autonomy model (L1--L4) is a novel attack surface unique to AI-driven CEA. An unprotected local REST endpoint allows escalation from L1 (monitoring only) to L4 (full autonomous optimization), enabling the AI to make irreversible control decisions without operator approval. This threat has no analogue in traditional ICS threat models. Mapped to T0855 (Unauthorized Command Message).

\subsection{Cloud and Integration Threats}

\textbf{T082 --- Multi-Tenant Data Leak via IDOR (I, DREAD: 42):} Insecure Direct Object Reference on the cloud dashboard allows enumeration of facility identifiers, exposing environmental telemetry, crop recipes, and yield data of competing tenants on the same platform. For cannabis operations, this data constitutes trade secrets worth \$500K--\$5M+. Mapped to T0811 (Data from Information Repositories).

\textbf{T093 --- Transfer Learning Model Injection (E, DREAD: 48):} An unauthenticated PUT to the cloud model repository allows an attacker to push arbitrary neural network weights to the fleet. Models are deserialized via frameworks (ONNX, PyTorch) that may execute arbitrary code during loading. A malicious model distributed across all downstream facilities constitutes a supply-chain attack with fleet-wide impact. Mapped to T0889 (Modify Program).

\textbf{T095 --- Compliance API Key Theft (S, DREAD: 46):} Stolen Metrc or BioTrack API credentials enable creation of fraudulent seed-to-sale tracking tags, triggering state-level regulatory investigations that can result in license suspension or revocation. Cannabis licenses represent \$250K--\$500K+ in direct cost and tens of millions in facility investment at risk. Mapped to T0859 (Valid Accounts).

\subsection{MITRE ATT\&CK for ICS Coverage}

The 123~threats map to 19~distinct MITRE ATT\&CK for ICS techniques across 10~of the 12~ICS tactics. The most frequently mapped techniques are T0814 (Denial of Service, 17~threats), T0872 (Indicator Removal on Host, 16~threats), T0848 (Rogue Master/Relay, 14~threats), and T0836 (Modify Parameter, 13~threats). The unmapped tactics are Collection and Command and Control, which are subsumed by the cloud-layer threat analysis.

\subsection{DREAD Severity Distribution}

\begin{table}[!t]
\centering
\caption{DREAD Severity Distribution}
\label{tab:dread_dist}
\begin{tabular}{@{}lcc@{}}
\toprule
\textbf{DREAD Range} & \textbf{Count} & \textbf{Interpretation} \\
\midrule
40--50 (Critical) & 47 & Immediate mitigation required \\
30--39 (High) & 48 & Scheduled mitigation \\
20--29 (Medium) & 27 & Backlog \\
$<$20 (Low) & 1 & Accept \\
\bottomrule
\end{tabular}
\end{table}

The ten highest-severity threats (DREAD~$\geq$~46) cluster around four categories: life-safety (CO\textsubscript{2} sensor manipulation T008, CO\textsubscript{2} injection override T055/T058), unauthorized command execution (T030, T048, T073, T074), supply chain compromise (T093, T101), and regulatory exposure (compliance API key theft T095). All ten are mitigable with existing technology; none require novel research.

\section{Physical Impact Analysis}
\label{sec:impact}

CEA cyberattacks differ fundamentally from traditional IT breaches because the target is a biological process with irreversible damage timelines. Table~\ref{tab:impact} quantifies the physical consequences of the six primary attack scenarios.

\begin{table*}[!t]
\centering
\caption{Physical Impact of Cyber-Physical Attacks on CEA Systems}
\label{tab:impact}
\begin{tabular}{@{}p{3.5cm}p{3.5cm}p{2.5cm}p{3cm}p{2.7cm}@{}}
\toprule
\textbf{Attack Scenario} & \textbf{Physical Consequence} & \textbf{Time to Damage} & \textbf{Financial Impact} & \textbf{Regulatory Impact} \\
\midrule
Temperature $+$5$^{\circ}$C above optimal & Pollen failure (tomato $>$32$^{\circ}$C), terpene loss (cannabis $>$30$^{\circ}$C), crop death at 45$^{\circ}$C & Hours to days & \$100K--\$1M+ per facility & None \\
\addlinespace
Humidity $>$85\% RH & Botrytis within 48--72\,h; powdery mildew within 3--7 days; VPD $<$0.4\,kPa & 48--72 hours & \$500K--\$1.5M (cannabis); remediation \$50K--\$200K & Failed state testing (TYM) \\
\addlinespace
CO\textsubscript{2} injection $>$5,000\,ppm & Worker headache/drowsiness; $>$40,000\,ppm: IDLH, loss of consciousness; $>$80,000\,ppm: death within minutes & Hours (sealed room with 50 SCFH system) & Safety incident + liability & OSHA citation \\
\addlinespace
Irrigation shutoff & Aeroponics: minutes; NFT: 2--4\,h; rockwool: 6--24\,h; coco: 12--48\,h & Minutes to days (substrate-dependent) & \$100K--\$500K & None \\
\addlinespace
Lighting dark-period violation & Cannabis hermaphroditism (90--95\% value destruction); flowering reversion & 2--3 nights & \$1.65M per 10,000\,sq\,ft event & None \\
\addlinespace
Ransomware (total lockout) & Cascading: temp rise (0--2\,h), irrigation failure (2--8\,h), mold germination (24--48\,h), total crop loss (72\,h--7\,d) & 72 hours to total loss & \$500K--\$5M+ & Compliance data loss \\
\bottomrule
\end{tabular}
\end{table*}

The cascading failure timeline under a full ransomware lockout is particularly severe: lights generate 30--50 BTU/sq~ft/hr, driving temperature increases within 2~hours; aeroponic and NFT systems begin failing within 2--4 hours as pump circulation stops; condensation and humidity buildup create pathogen conditions within 24--48 hours; and total crop loss occurs within 72~hours to 7~days depending on substrate type and season.

For cannabis specifically, the financial exposure is disproportionate to facility size. Indoor cannabis flower yields 35--70\,g/sq~ft per harvest cycle at wholesale prices of \$1,000--3,100/lb, producing \$165/sq~ft per cycle. A single 10,000~sq~ft flowering room represents approximately \$1.65~million per harvest, and a hermaphroditism event from lighting disruption can destroy 90--95\% of this value in a single incident.

\section{Adversarial Machine Learning Threats}
\label{sec:ai_threats}

Modern CEA platforms incorporate machine learning components that introduce a novel attack surface not covered by traditional ICS threat models. Our literature review confirms that \textbf{no published work addresses adversarial ML attacks specifically against CEA control systems}. We identify five novel attack classes by composing established adversarial ML techniques with CEA-specific physics, biology, and economics.

\subsection{Stealth Destabilization of NN-Tuned PID (Novel)}

The 7-3-3 MLP that auto-tunes PID gains ($K_p$, $K_i$, $K_d$) operates as an online learning system. An attacker with sensor-channel access injects slow, low-amplitude biased measurements over days or weeks. The MLP accumulates toward a local optimum that produces marginally stable gains. Control appears correct until a disturbance (door opening, lighting transition) triggers oscillation that overshoots setpoints enough to damage flowering-stage crops.

Unlike classic PID attacks (e.g., Stuxnet~\cite{ref_stuxnet}, which manipulated controller \textit{outputs}), this attack manipulates the \textit{tuner} --- leaving a much smaller forensic footprint attributable to ``model drift'' rather than compromise. The attack is feasible because the Jagielski~\textit{et~al.}~\cite{ref_jagielski} poisoning framework for regression models applies directly to the 7-3-3 MLP's continuous $K_p/K_i/K_d$ output space.

\subsection{Baseline Drift Poisoning of Anomaly Detectors (Novel)}

The autoencoder-based anomaly detector retrains on rolling windows to accommodate legitimate drift from plant growth stages and seasonal HVAC behavior. An attacker gradually expands the ``normal'' manifold by injecting small, slowly drifting anomalies during retraining windows. After $N$~weeks, the detector accepts clearly abnormal conditions as normal, enabling a subsequent payload attack (e.g., humidity spike) to proceed undetected.

This attack extends the constrained concealment framework of Erba~\textit{et~al.}~\cite{ref_erba} from water treatment to CEA, exploiting the fact that \textit{legitimate} concept drift in CEA provides cover for \textit{malicious} drift.

\subsection{Cross-Facility Transfer Learning Propagation (Novel)}

Cross-facility transfer learning is a core CEA value proposition: models learned at Facility~A are deployed to Facility~B without sharing proprietary data. A compromised facility (via employee, supply chain, or intrusion) pushes a backdoored weight delta into the global model via the federated update channel. The backdoor is triggered by a specific environmental pattern (e.g., CO\textsubscript{2}$=$842~ppm $\wedge$ EC$=$2.31~mS/cm) unlikely to occur naturally but inducible by the attacker at the target facility later.

This attack composes Bagdasaryan~\textit{et~al.}'s~\cite{ref_bagdasaryan} federated learning backdoor with Turner~\textit{et~al.}'s clean-label technique~\cite{ref_turner_clean_label}. The low participant count in CEA federations (5--30~facilities vs.\ millions of mobile devices) dramatically increases per-attacker influence, and no CEA platform deploys Byzantine-robust aggregation (Krum, trimmed mean) --- naive FedAvg is the norm.

\subsection{Adversarial Agronomic Schedules (Novel)}

This is the most novel and CEA-unique attack class. An adversary generates irrigation, fertigation, or lighting schedules that are:
\begin{itemize}\item In-distribution for the anomaly detector (no alarm triggered)
\item In-spec for regulatory compliance limits
\item Agronomically sound on paper (individual parameters within accepted ranges)
\end{itemize}

\noindent Yet these schedules interact with crop biology --- specific cultivar $\times$ growth stage $\times$ historical stress --- to cause physiological damage: tip-burn, blossom-end rot, nutrient lockout, or flowering reversion.

This is the first reported adversarial ML attack class where the ``classifier'' being fooled is a \textbf{living organism} rather than a computational model. The perturbation budget is defined by agronomic norms rather than $L_p$ balls, and the defender's anomaly detector sees nothing wrong because nothing \textit{is} wrong numerically --- the crop nonetheless fails. This attack requires horticultural domain knowledge, but this knowledge is available in the public literature for every major CEA crop.

\subsection{Reward Poisoning of RL Energy Optimizer (Novel)}

The cascading VPD energy optimizer uses reinforcement learning with a multi-objective reward function combining energy cost minimization and environmental stability. Reward poisoning~\cite{ref_ma_policy_poison} biases the optimizer toward marginal setpoints that save energy measurably but degrade crop quality subtly (e.g., slightly lower night VPD that promotes pathogen growth while improving kWh/kg on the attacker's chosen metric). The KPI looks better; the crop looks worse over months. Detection requires ground-truth crop quality audits --- a feedback loop measured in biological time (weeks), not computational time (seconds).

\section{CEA Vendor Security Survey}
\label{sec:vendors}

We surveyed the cybersecurity posture of 10~commercial CEA control system vendors. Table~\ref{tab:vendors} summarizes the findings.

\begin{table*}[!t]
\centering
\caption{CEA Control System Vendor Cybersecurity Posture}
\label{tab:vendors}
\begin{tabular}{@{}llllcccc@{}}
\toprule
\textbf{Vendor} & \textbf{HQ} & \textbf{Segment} & \textbf{Protocols} & \textbf{CVEs} & \textbf{Bug Bounty} & \textbf{62443 Cert} & \textbf{SOC~2} \\
\midrule
Priva & Netherlands & Greenhouse & BACnet, Modbus & 1 & No & No & No \\
Argus Controls & Canada & Research/Cannabis & Proprietary $\rightarrow$ BACnet/Modbus & 0 & No & No & No \\
TrolMaster & China & Cannabis & Proprietary, AWS cloud & 0 & No & No & No \\
Wadsworth & USA & Greenhouse & BACnet & 0 & No & No & No \\
Hoogendoorn & Netherlands & Greenhouse & Proprietary, BACnet & 0 & No & No & No \\
Ridder/HortiMaX & Netherlands & Greenhouse & BACnet, Modbus & 0 & No & No & No \\
Growlink & USA & Cannabis & BACnet, Modbus, REST API & 0 & No & No & No \\
Autogrow (Priva) & New Zealand & Indoor/Cannabis & NATS, REST API & 0 & No & No & No \\
iUNU/LUNA & USA & Vision/Monitoring & Cloud-only (no actuator control) & 0 & No & No & No \\
Source.ag & Netherlands & AI Overlay & Reads/writes to underlying BMS & 0 & No & No & No \\
\bottomrule
\end{tabular}
\end{table*}

The single CVE (CVE-2022-3010) affects Priva's TopControl Suite~\cite{ref_priva_cve}: a weak password hash (CWE-916) with CVSS~7.5, disclosed by NorthWave via the Dutch Institute for Vulnerability Disclosure (DIVD). No other CEA vendor has received a CISA ICS-CERT advisory, conducted a public penetration test, or published a security architecture document.

This finding is striking when compared to adjacent sectors. Building automation vendors (Honeywell/Tridium, Johnson Controls, Schneider Electric) have accumulated hundreds of CVEs, operate coordinated disclosure programs, and increasingly pursue IEC~62443 certification. The CEA vendor category is a decade behind in cybersecurity maturity.

Supply chain jurisdiction is a concern: TrolMaster, the dominant vendor in U.S. cannabis cultivation, is designed and manufactured in China (Xiamen). Under the PRC National Intelligence Law (2017) and Data Security Law (2021), Chinese firms may be legally compelled to assist state intelligence operations. TrolMaster's cloud tunnel is always-on by default, routing all facility telemetry through AWS endpoints configured by the vendor. The firmware is closed-source with no published security audit.

\section{Countermeasure Framework}
\label{sec:countermeasures}

We propose a defense-in-depth framework organized by IEC~62443 security levels, with countermeasures mapped to IEC~62443-3-3 foundational requirements (FR), NIST CSF v2.0 functions, and OWASP IoT Top~10~\cite{ref_owasp_iot} categories.

\subsection{Network Segmentation (FR~5: Restricted Data Flow)}

CEA facilities should implement the Purdue Enterprise Reference Architecture with at least four VLANs: OT-Control (PLCs, RTUs), OT-Sensors, IT-Corporate, and a DMZ between OT and IT networks. For SME farms with budgets under \$10K, cost-effective implementation uses pfSense/OPNsense open-source firewalls (\$300--500 hardware), managed switches with VLAN support (Cisco CBS250, Ubiquiti USW-Pro-24, \$200--600), and SecurityOnion for SIEM/IDS integration (free). Enterprise deployments should consider unidirectional gateways (data diodes) from Waterfall Security or Owl Cyber Defense for the OT-to-IT boundary.

\subsection{Protocol Security (FR~4: Data Confidentiality)}

BACnet Secure Connect (BACnet/SC, ASHRAE~135 Addendum~bj) adds TLS~1.3 and X.509 certificate-based device authentication but has seen minimal adoption in CEA-specific controllers. The Modbus/TCP Security Specification (2018) defines TLS wrapping on port~802 but is effectively unadopted. For practical deployment, we recommend: (1)~MQTT with TLS and client certificates as the primary IoT telemetry protocol; (2)~WireGuard VPN tunnels for Modbus/TCP segments crossing untrusted networks; (3)~migration from analog I/O (4--20\,mA, 0--10\,V) to IO-Link (IEC~61131-9) for new installations; and (4)~physical protection (metallic conduit, tamper-evident enclosures) for legacy analog wiring.

\subsection{Monitoring and Detection (FR~6: Timely Response)}

We recommend a dual detection strategy combining network anomaly detection (Suricata~7.x with ICS rulesets on a passive SPAN tap) and process anomaly detection (statistical process control at minimum; autoencoder-based ML detection for high-value zones). Process anomaly detection is more valuable for CEA than network anomaly detection because attacks via compromised legitimate devices produce normal-looking network traffic while creating abnormal process behavior.

A CEA-unique defense opportunity exists: \textbf{crop growth rate as an integrity signal}. Computer vision monitoring of plant height, leaf area index, and canopy color provides an independent verification channel. If sensors report optimal conditions but visual crop health diverges from the growth model, the sensors may be compromised. This defense is unavailable in any other ICS domain.

\subsection{AI/ML Security}

Defenses against the five novel AI/ML threats (Section~\ref{sec:ai_threats}) require:
\begin{itemize}\item Gain rate-of-change limiters enforced at the actuator driver level, independent of the NN tuner
\item Out-of-band ``golden baseline'' anomaly models retained from commissioning, independent of rolling retraining
\item Byzantine-robust aggregation (Krum~\cite{ref_krum}, FLTrust~\cite{ref_fltrust}) for cross-facility transfer learning
\item Plant-response-aware anomaly detection over schedule trajectories, not just point values
\item Multi-objective RL reward functions with crop-quality constraints verified by ground-truth audits
\end{itemize}

\subsection{Incident Response}

CEA incident response differs critically from traditional IT: \textbf{do not isolate OT systems by shutting them down}. A greenhouse without climate control in summer can reach lethal temperatures within one hour. Instead, sever external access at the DMZ firewall while maintaining internal control loops, switch affected systems to manual/local mode, and deploy personnel with handheld instruments (thermometer, hygrometer, pH pen, EC pen, CO\textsubscript{2} monitor) for physical monitoring.

\subsection{Recommended Security Levels}

Based on our threat analysis, we recommend:
\begin{itemize}\item \textbf{SL~1 (Basic hygiene)}: all CEA operations --- change default credentials, segment networks, patch, maintain offline backups.
\item \textbf{SL~2 (Managed)}: commercial CEA --- dedicated firewalls, IDS, RBAC, encrypted protocols, incident response plan.
\item \textbf{SL~3 (Proactive)}: high-value/regulated operations (pharmaceutical cannabis, research facilities) --- certificate-based device authentication, ML anomaly detection, penetration testing, SBOM management.
\end{itemize}

\section{Discussion}
\label{sec:discussion}

\subsection{Comparison with Adjacent Domains}

Our enumeration of 123~threats compares favorably with adjacent threat models: Fereidooni~\textit{et~al.}~\cite{ref_fereidooni} identified 58~threats for precision agriculture (we cover a broader stack and find 2.1$\times$ more), and Tripathi~\textit{et~al.}~\cite{ref_tripathi} identified 126~threats for smart greenhouses (comparable count but our threats include cloud, ML, and compliance layers they did not address). The building automation security analysis~\cite{ref_bas_security} identifies protocol-level BACnet and Modbus vulnerabilities that transfer directly to CEA but does not address the agricultural impact layer.

\subsection{Novel Contributions Beyond Traditional ICS}

Three aspects of CEA create threat categories absent from general ICS threat models:

\textbf{Biological targets}: CEA is the only ICS domain where the process output is a living organism with irreversible damage timelines measured in hours. This creates asymmetric risk --- a 4-hour HVAC outage in a commercial building is an inconvenience; in a flowering cannabis facility, it is a \$1.65M loss.

\textbf{Progressive autonomy}: The L1--L4 autonomy model introduces a novel privilege escalation vector (T066) with no analogue in traditional SCADA.

\textbf{Federated ML}: Cross-facility transfer learning creates multi-tenant poisoning vectors (T090--T093) not addressed by IEC~62443 or NIST SP~800-82.

\textbf{Safety-security coupling}: CEA facilities with CO\textsubscript{2} enrichment systems operate safety-critical interlocks (high-CO\textsubscript{2} alarms, emergency exhaust fans) that can be disabled through the same unauthenticated OT protocols used for normal control. The TRITON/TRISIS precedent~\cite{ref_triton} demonstrated that safety systems are not immune to targeted compromise when they share network infrastructure with control systems. ISA TR84.00.09 (Cybersecurity Related to the Safety Lifecycle) provides guidance for integrating cybersecurity into Safety Instrumented System design, but no CEA vendor or operator is known to have adopted it.

\textbf{Legacy and vendor-failure risk}: The CEA industry faces a growing population of stranded control systems from vendor failures. Between 2022 and 2023, five major indoor farming operators --- InFarm, AeroFarms (Chapter~11), Fifth Season, Kalera, and AppHarvest --- declared bankruptcy~\cite{ref_indoor_farm_failures}, each leaving behind custom control stacks with no ongoing security support. Hoogendoorn's flagship iSii process computer reached end-of-sale on January~1, 2026, creating a population of Linux-based controllers that will operate for 15--20 years with declining patch cadence. These legacy systems represent an expanding attack surface with no responsible party for vulnerability management.

\subsection{Regulatory Implications}

The U.S. regulatory vacuum for CEA cybersecurity --- voluntary CISA guidelines with no mandatory requirements, no FSMA cybersecurity component, no state-level cannabis cybersecurity mandates --- stands in contrast to the EU NIS2 Directive, which explicitly includes food production as an ``important entity'' subject to mandatory risk management. We argue that CEA systems managing controlled substances (cannabis), food crops under FSMA jurisdiction, or facility environments with worker-safety-critical CO\textsubscript{2} systems should be subject to mandatory cybersecurity baselines aligned with IEC~62443 SL~2.

\subsection{Limitations}

This threat model is developed against a single vendor's architecture. While the three-tier pattern (field/edge/cloud) and the protocol portfolio (Modbus, BACnet, MQTT) are representative of the broader CEA industry, vendor-specific implementation details (e.g., TrolMaster's proprietary bus protocol, Priva's Windows-based supervisory layer) introduce threats not captured here. The DREAD scoring methodology, while practical, is subjective; future work should apply FAIR quantitative risk analysis with Monte Carlo simulation. The expert validation (Delphi) planned for this work was not completed within the submission timeline and will be reported in a follow-up study.

\subsection{Ethical Considerations}

This paper discloses no zero-day vulnerabilities. All threats are derived from publicly documented protocol weaknesses, published CVEs, and architectural analysis. The decision to publish aligns with the Kohnfelder and Shostack~\cite{ref_publish_tm} position that published threat models improve collective defense. Specific facility locations, IP addresses, and customer identities are not disclosed.

\section{Conclusion}
\label{sec:conclusion}

This paper presents the first comprehensive threat model for IoT-enabled Controlled Environment Agriculture systems. Systematic STRIDE analysis of a production platform deployed across 30+ facilities yields 123~unique threats across 25~DFD elements and 15~communication protocols, mapped to 19~MITRE ATT\&CK for ICS techniques and scored using DREAD risk assessment. We identify five novel attack classes unique to AI-driven CEA, including adversarial agronomic schedules --- the first adversarial ML attack targeting a biological organism rather than a computational model.

The near-total absence of public cybersecurity posture among CEA vendors (one CVE across ten vendors, zero bug bounties, zero IEC~62443 certifications) represents a systemic risk to a critical infrastructure sector. The U.S. regulatory vacuum --- no mandatory cybersecurity requirements despite critical infrastructure designation --- leaves CEA operators without compliance incentives or guidance.

We recommend that: (1)~CEA operators adopt IEC~62443 SL~2 as a minimum baseline; (2)~regulatory bodies extend mandatory cybersecurity requirements to CEA facilities managing food crops, controlled substances, or worker-safety-critical systems; (3)~CEA vendors establish coordinated vulnerability disclosure programs and pursue IEC~62443-4-1 product development certification; and (4)~the research community develop CEA-specific digital twin cyber ranges to enable safe security testing without risking production crops.

The complete threat catalog, DFD artifacts, and DREAD scoring matrices are provided as supplementary materials accompanying this paper.


\newcommand{\authorphoto}{\includegraphics[width=1in,height=1.25in,clip,keepaspectratio]{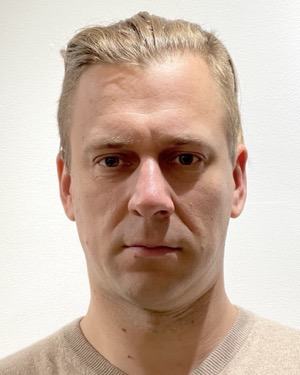}}
\begin{IEEEbiography}[\authorphoto]{Andrii Vakhnovskyi}
received the B.S. degree in computer engineering and the M.S. degree in systems engineering from the National Technical University ``Kharkiv Polytechnic Institute'' (NTU ``KhPI''), Ukraine, in 2009 and 2011, respectively. He is the Founder and CEO of IOGRU LLC, New York, NY, where he develops AI-driven IoT platforms for climate control in controlled environment agriculture. His systems have been deployed across 30+ commercial facilities in 8 U.S. climate zones, managing over 10 million square feet of cultivation space. He is a Senior Member of ISA and a Member of IEEE. His research interests include industrial IoT security, neural network control, and adversarial machine learning in cyber-physical systems.
\end{IEEEbiography}


\begin{thebibliography}{99}

\bibitem{ref_cea_market}
Research Nester, ``Controlled environment agriculture market size, share \& trends analysis report,'' 2025. [Online]. Available: https://www.researchnester.com/reports/controlled-environment-agriculture-market/6650

\bibitem{ref_iogru_arxiv}
A.~Vakhnovskyi, ``IOGRUCloud: A scalable AI-driven IoT platform for climate control in controlled environment agriculture,'' \textit{arXiv preprint arXiv:2604.07586}, 2026.

\bibitem{ref_cisa_sector}
Cybersecurity and Infrastructure Security Agency, ``Food and agriculture sector,'' 2024. [Online]. Available: https://www.cisa.gov/topics/critical-infrastructure-security-and-resilience/critical-infrastructure-sectors/food-and-agriculture-sector

\bibitem{ref_fbi_pin_2022}
Federal Bureau of Investigation, ``Ransomware attacks on agricultural cooperatives potentially timed to critical seasons,'' Private Industry Notification, Apr. 2022.

\bibitem{ref_halcyon_2025}
Halcyon, ``Ransomware attacks targeting agriculture and food production doubled in 2025,'' Halcyon Blog, 2025.

\bibitem{ref_jbs}
B.~Fung, ``JBS paid \$11 million to resolve ransomware attack,'' \textit{CNN Business}, Jun. 2021. [Online]. Available: https://www.cnn.com/2021/06/09/business/jbs-cyberattack-ransom-paid/

\bibitem{ref_new_coop}
J.~Greig, ``BlackMatter ransomware hits Iowa grain cooperative NEW Cooperative,'' \textit{ZDNet}, Sep. 2021. [Online]. Available: https://www.zdnet.com/article/blackmatter-ransomware-hits-iowa-grain-cooperative/

\bibitem{ref_stiiizy}
L.~Abrams, ``STIIIZY data breach exposes cannabis buyers' IDs and purchases,'' \textit{BleepingComputer}, Jan. 2025. [Online]. Available: https://www.bleepingcomputer.com/news/security/stiiizy-data-breach/

\bibitem{ref_jbair_microgrid}
M.~Jbair, B.~Ahmad, and R.~Harrison, ``Threat modelling of cyber-physical systems --- a case study of a microgrid system,'' \textit{Computers \& Security}, vol.~124, 2023.

\bibitem{ref_automotive_tara}
R.~Moreira, E.~Cust\'odio, and A.~Pinto, ``A systematic review of TARA methodologies for connected and automated vehicles,'' \textit{IEEE Access}, vol.~12, pp.~42560--42583, 2024.

\bibitem{ref_healthcare_stride}
M.~Z.~Hasan, R.~Hasan, and S.~Islam, ``STRIDE-based threat modeling and risk assessment framework for IoT-enabled smart healthcare systems,'' \textit{Sensors}, vol.~25, no.~3, 2025.

\bibitem{ref_bas_security}
R.~Kaur, D.~Gabrijelcic, and T.~Peceny, ``On building automation system security,'' \textit{Internet of Things}, vol.~25, p.~101063, Elsevier, 2024.

\bibitem{ref_fereidooni}
H.~Fereidooni, A.~Taheri, and A.-R.~Sadeghi, ``STRIDE-based cyber security threat modeling for IoT-enabled precision agriculture systems,'' in \textit{Proc. IEEE CCNC}, 2022, pp.~955--960. DOI: 10.1109/CCNC49032.2022.9732597.

\bibitem{ref_tripathi}
N.~Tripathi, N.~Hubballi, and Y.~Singh, ``A study on threat modeling in smart greenhouses,'' \textit{J. Inform. Security Cybercrimes Res.}, 2021.

\bibitem{ref_shostack}
A.~Shostack, \textit{Threat Modeling: Designing for Security}. Wiley, 2014.

\bibitem{ref_attack_ics}
MITRE, ``ATT\&CK for Industrial Control Systems,'' 2020. [Online]. Available: https://attack.mitre.org/matrices/ics/

\bibitem{ref_iec62443}
ISA/IEC 62443, ``Industrial automation and control systems security,'' International Society of Automation, 2013--2024.

\bibitem{ref_dread}
M.~Howard and D.~LeBlanc, \textit{Writing Secure Code}, 2nd~ed. Microsoft Press, 2002.

\bibitem{ref_xiong_slr}
W.~Xiong and R.~Lagerstr\"om, ``Threat modeling --- a systematic literature review,'' \textit{Computers \& Security}, vol.~84, pp.~53--69, 2019.

\bibitem{ref_shevchenko_survey}
N.~Shevchenko \textit{et~al.}, ``Threat modeling: A summary of available methods,'' SEI CMU, Tech. Rep., 2018.

\bibitem{ref_publish_tm}
D.~Kohnfelder and A.~Shostack, ``Publish your threat models!'' \textit{arXiv preprint arXiv:2511.08295}, 2025.

\bibitem{ref_ferrag_survey}
M.~A.~Ferrag \textit{et~al.}, ``Deep learning for cyber security intrusion detection: Approaches, datasets, and comparative study,'' \textit{J. Inform. Security Appl.}, 2020.

\bibitem{ref_gupta_review}
M.~Gupta \textit{et~al.}, ``A review on security of smart farming and precision agriculture,'' \textit{Applied Sciences}, vol.~11, no.~16, 2021.

\bibitem{ref_cea_cyber_2024}
A.~Alahmadi, N.~Alkhatib, and M.~Alardhi, ``Cyber security in smart agriculture: Threat types, current status, and future trends,'' \textit{Computers and Electronics in Agriculture}, vol.~224, p.~109202, 2024.

\bibitem{ref_cs_slr_2024}
M.~Hossain, Y.~Sani, and S.~Kashem, ``Cybersecurity in smart agriculture: A systematic literature review,'' \textit{Computers \& Security}, vol.~146, p.~104051, 2024.

\bibitem{ref_kulkarni_incidents}
S.~Kulkarni \textit{et~al.}, ``A review of cybersecurity incidents in the food and agriculture sector,'' \textit{Smart Agricultural Technology}, 2025. arXiv:2403.08036.

\bibitem{ref_murch_2018}
R.~S.~Murch \textit{et~al.}, ``Cyberbiosecurity: An emerging new discipline to help safeguard the bioeconomy,'' \textit{Frontiers in Bioengineering and Biotechnology}, 2018.

\bibitem{ref_duncan_2019}
S.~E.~Duncan \textit{et~al.}, ``Cyberbiosecurity: A new perspective on protecting U.S. food and agricultural system,'' \textit{Frontiers in Bioengineering and Biotechnology}, vol.~7, p.~63, 2019.

\bibitem{ref_ics_tm_slr}
A.~Humayed, J.~Lin, F.~Li, and B.~Luo, ``Threat modeling of industrial control systems: A systematic literature review,'' \textit{Computers \& Security}, vol.~137, p.~103617, 2024.

\bibitem{ref_food_ag_isac}
Food and Ag-ISAC, ``72 active threat actors targeting food supply chains,'' \textit{Industrial Cyber}, 2025.

\bibitem{ref_hunt_hackett}
Hunt \& Hackett, ``Agriculture in the crosshairs of nation-state sponsored hackers,'' 2024.

\bibitem{ref_stuxnet}
R.~Langner, ``Stuxnet: Dissecting a cyberwarfare weapon,'' \textit{IEEE Security \& Privacy}, vol.~9, no.~3, pp.~49--51, 2011.

\bibitem{ref_jagielski}
M.~Jagielski, A.~Oprea, B.~Biggio, C.~Liu, C.~Nita-Rotaru, and B.~Li, ``Manipulating machine learning: Poisoning attacks and countermeasures for regression learning,'' in \textit{Proc. IEEE S\&P}, 2018.

\bibitem{ref_erba}
A.~Erba \textit{et~al.}, ``Constrained concealment attacks against reconstruction-based anomaly detectors in industrial control systems,'' in \textit{Proc. ACSAC}, 2020.

\bibitem{ref_bagdasaryan}
E.~Bagdasaryan, A.~Veit, Y.~Hua, D.~Estrin, and V.~Shmatikov, ``How to backdoor federated learning,'' in \textit{Proc. AISTATS}, 2020.

\bibitem{ref_turner_clean_label}
A.~Turner, D.~Tsipras, and A.~Madry, ``Clean-label backdoor attacks,'' in \textit{ICLR Workshop}, 2019.

\bibitem{ref_ma_policy_poison}
Y.~Ma, X.~Zhang, W.~Sun, and J.~Zhu, ``Policy poisoning in batch reinforcement learning and control,'' in \textit{Proc. NeurIPS}, 2019.

\bibitem{ref_krum}
P.~Blanchard, E.~M.~El~Mhamdi, R.~Guerraoui, and J.~Stainer, ``Machine learning with adversaries: Byzantine tolerant gradient descent,'' in \textit{Proc. NeurIPS}, 2017.

\bibitem{ref_fltrust}
X.~Cao \textit{et~al.}, ``FLTrust: Byzantine-robust federated learning via trust bootstrapping,'' in \textit{Proc. NDSS}, 2021.

\bibitem{ref_nist_800_82}
NIST, ``Guide to operational technology (OT) security,'' NIST SP 800-82 Rev.~3, Sep. 2023.

\bibitem{ref_owasp_iot}
OWASP, ``IoT Top 10,'' 2018. [Online]. Available: https://owasp.org/www-project-internet-of-things/

\bibitem{ref_priva_cve}
CISA, ``Priva TopControl Suite,'' ICSA-22-356-01, Dec. 2022. CVE-2022-3010, CVSS 7.5.

\bibitem{ref_triton}
A.~Di~Pinto, Y.~Dragoni, and A.~Carcano, ``TRITON: How it disrupted safety systems and changed the threat landscape of industrial control systems forever,'' in \textit{Proc. Black Hat USA}, 2018.

\bibitem{ref_blackenergy}
R.~M.~Lee, M.~J.~Assante, and T.~Conway, ``Analysis of the cyber attack on the Ukrainian power grid,'' Electricity Information Sharing and Analysis Center (E-ISAC) and SANS ICS, Mar. 2016.

\bibitem{ref_niagara_cves}
CISA, ``Honeywell/Tridium Niagara Framework multiple vulnerabilities,'' ICS-CERT Advisories, 2025. [13~CVEs disclosed in 2025 affecting Niagara~4 Framework versions prior to 4.14.]

\bibitem{ref_basc20t_cve}
CISA, ``Contemporary Controls BAScontrol BASC-20T unauthenticated remote code execution,'' ICS-CERT Advisory, CVE-2025-13926, 2025.

\bibitem{ref_mo_hailong}
U.S. Department of Justice, ``Chinese citizen sentenced on charges of conspiring to steal trade secrets,'' Press Release, Oct. 2016. [Online]. Available: https://www.justice.gov/opa/pr/chinese-citizen-sentenced-charges-conspiring-steal-trade-secrets

\bibitem{ref_haitao_xiang}
U.S. Department of Justice, ``Former Monsanto scientist sentenced for stealing trade secrets,'' Press Release, Nov. 2017.

\bibitem{ref_indoor_farm_failures}
H.~Pham, ``Indoor farming's reckoning: AeroFarms, AppHarvest, and the vertical farming shakeout,'' \textit{AgFunderNews}, 2023. [Online]. Available: https://agfundernews.com/indoor-farming-shakeout

\end{thebibliography}
\end{document}